\documentclass[aip, apl, twocolumn, superscriptaddress, preprintnumbers, amsmath, pdfFig, floatfix,normalem]{revtex4}
\usepackage{graphicx}
\usepackage{dcolumn}
\usepackage{bm}
\usepackage{ulem}
\usepackage{amsmath}
\usepackage{textcomp}
\usepackage{commath}
\usepackage{mathtools}
\usepackage{graphicx}
\usepackage{bm}
\usepackage{braket}
\usepackage{esvect}

\usepackage[utf8]{inputenc}
\usepackage[english]{babel}
\usepackage{keyval}
\graphicspath{{Figs/}{Figs:}{\Figs}}
\setkeys{Gin}{width=0.9\columnwidth}

\usepackage[usenames,dvipsnames]{color}
\usepackage[]{ulem}

\newcommand{\comment}[1]{\textcolor{red}{#1}}
\renewcommand{\comment}[1]{\relax}

\newcommand{\todelete}[1]{\textcolor{green}{\sout{#1}}}
\renewcommand{\todelete}[1]{\relax}

\begin{document}

\title{Planar Hall Effect and Anisotropic Magnetoresistance in a polar-polar interface of LaVO$_3$-KTaO$_3$ with strong spin-orbit coupling  }
\date{\today}
\author{Neha Wadehra}
\affiliation{Nanoscale Physics and Device Laboratory, Institute of Nano Science and Technology, Phase-10, Sector-64, Mohali, Punjab, 160062, India.}

\author{Ruchi Tomar}
\affiliation{Nanoscale Physics and Device Laboratory, Institute of Nano Science and Technology, Phase-10, Sector-64, Mohali, Punjab, 160062, India.}

\author{R.K Gopal}
\affiliation{Indian Institute of Science Education and Research Mohali, Knowledge City, Sector-81, 
SAS Nagar,  Manauli, 140306, India.}

\author{Yogesh Singh}
\affiliation{Indian Institute of Science Education and Research Mohali, Knowledge City, Sector-81, 
SAS Nagar,  Manauli, 140306, India.}

\author{Sushanta Dattagupta}
\affiliation{Bose Institute, P-1/12, CIT Rd, Scheme VIIM, Kankurgachi, Kolkata, West Bengal-700054, India.}

\author{S. Chakraverty}
\email{suvankar.chakraverty@gmail.com }
\affiliation{Nanoscale Physics and Device Laboratory, Institute of Nano Science and Technology, Phase-10, Sector-64, Mohali, Punjab, 160062, India.}

\begin{abstract}
\noindent
Among the perovskite oxide family, KTaO$_3$ (KTO) has recently attracted considerable interest as a possible system for the realization of the Rashba effect. In this work, we improvise a novel conducting interface by juxtaposing KTO with another insulator, namely LaVO$_3$ (LVO) and report planar Hall effect (PHE) and anisotropic magnetoresistance (AMR) measurements. This interface exhibits a signature of strong spin-orbit coupling. Our experimental observation of two fold AMR at low magnetic fields can be intuitively understood using a phenomenological theory for a Rashba spin-split system. At high fields ($\sim$8 T), we see a two fold to four fold transition in the AMR that could not be explained using only Rashba spin-split energy spectra. We speculate that it might be generated through an intricate process arising from the interplay between strong spin-orbit coupling, broken inversion symmetery, relativistic conduction electron and possible uncompensated localized vanadium spins.
\end{abstract}

\maketitle


In recent times, the urge of attaining new functionalities in modern electronic devices has led to the manipulation of spin degree of freedom of an electron along with its charge. \cite{Datta1990,Meier2002} This has given rise to an altogether new field of spin-electronics or "spintronics". It has been realized that momentum dependent splitting of spin-bands in an electronic system, the "Rashba effect", might play a key role in spintronic devices. \cite{Bychkov1984} The Rashba effect is important not only because it might have tremendous potential for technical applications, but also because it is a hunting ground of emergent physical properties.\cite{Inoue2009,Murakawa2013}

Semiconducting materials such as heterostructures of GaAs/GaAlAs and InAs/InGaAs have already been explored for the manifestation of the Rashba effect.\cite{Eisentein1984,Nitta1997} Another potentially rewarding class of materials for realization of this effect is "oxides".\cite{Bibes2011,Shanavas2014} The benefit of using oxides for spin based electronic devices is that they manifest a wealth of functional properties like magnetoresistance, superconductivity, ferromagnetism, ferroelectricity, charge ordering etc. which can be coupled with the Rashba effect to achieve emergent phenomena if a suitable interface or superlattice is designed.\cite{Tokura1994,Schooley1964,Tikhomirov2002,Lee2013} In addition to this, simple cubic structure of perovskite oxides makes them easily usable for fabrication of heterostructures for device applications.\cite{Bibes2011}

Among perovskite oxides, SrTiO$_3$ (STO) has been widely explored for realization of 2DEG at its interface with other perovskite oxides such as LaAlO$_3$ (LAO), LaVO$_3$ (LVO) and CaZrO$_3$ (CZO) etc.\cite{Ohtomo2004,Hotta2007,Chen2015} However, the spin orbit coupling strength of STO (which is a prerequisite for realization of the Rashba effect) is not very high. Another promising candidate from the perovskite oxide family having potential to host low dimensional electron gas is KTO.\cite{Tomar2017,Shanavas2014a} This insulating material has a dielectric constant and band gap similar to STO with an additional advantage of having strong spin orbit coupling (SOC) strength due to presence of 5d Tantalum atoms.\cite{Wadehra2017} The energy level splitting in KTO due to SOC is around 400 meV which is an order of magnitude higher than that of STO (17 meV).\cite{Nakamura2009}

With the aim of realizing 2DEG in a perovskite oxide with strong spin-orbit coupling, we grew heterostructure between LVO and KTO. The heterointerface was found to be conducting above the film thickness of 3 monolayers (ml). A  carrier mobility of around 600 cm$^2$V$^{-1}$s$^{-1}$ was measured at the interface with varying thickness of the LVO film. Anisotropic magnetoresistance which is a relativistic magnetotransport phenomenon observed in magnetic and some topological systems is predicted in systems with the Rashba-Dresselhaus type spin-splitting.\cite{Ky1968,Goennenwein2007,Taskin2017,Rakhmilevich2018,Trushin2009,Kozlov2019} We also report the observation of planar Hall effect and oscillations in longitudinal anisotropic magnetoresistance in our LVO-KTO system. A theoretical modelling using Rashba spin-split energy spectrum could predict our observation of 2 fold oscillations in AMR at low applied magnetic fields. The appearance of an additional periodicity in AMR above 8 T magnetic field suggests a possible complex and rich physics arising from the interplay between uncompensated localized vanadium spins, relativistic 2 dimensional itinerant electrons and strong spin-orbit coupling present in the system. 

\begin{figure}[t!] \scalebox{0.8}{\includegraphics{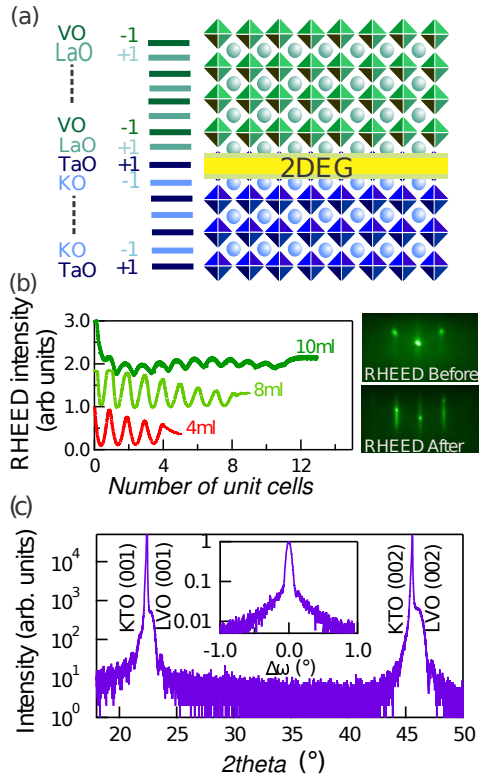}}
\caption{(Color online) (a) Schematic of the LVO-KTO heterostructure showing alternately charged layeres in both LVO and KTO leading to formation of 2DEG at the interface (b) RHEED oscillations for 4, 8 and 10 ml LVO-KTO samples. (right panel) RHEED pattern for 10 ml sample before and after growth of LVO film. (c) X-ray diffraction pattern of 40 ml sample showing crystalline film growth of LVO on KTO. }
\end{figure}

\begin{figure}[t!] \scalebox{0.8}{\includegraphics{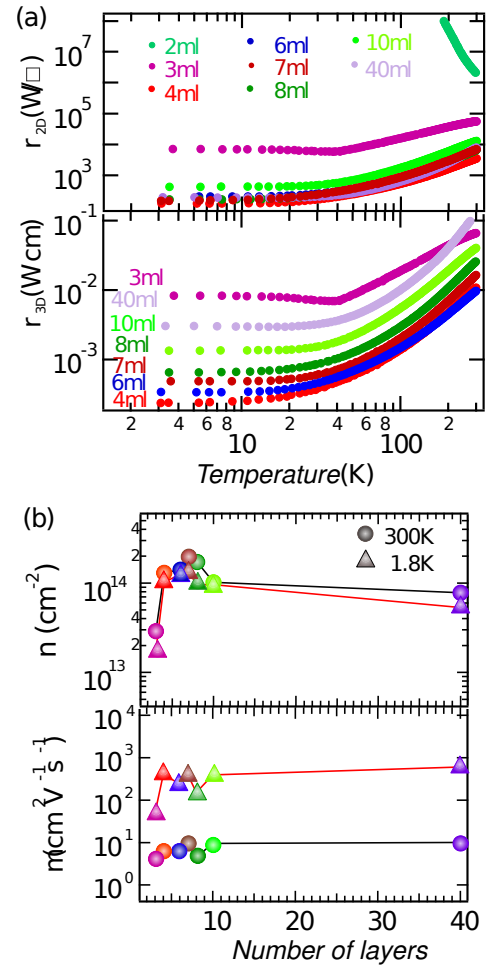}}
\caption{(Color online) (a) Temperature dependent 2D resistivity (upper panel) and 3D resistivity (lower panel) for LVO-KTO samples with varying LVO thickness. (b) charge carrier density and mobility of all the samples measured at 300 K and 1.8 K.}
\end{figure}

\begin{figure}[t!] \scalebox{1.1}{\includegraphics{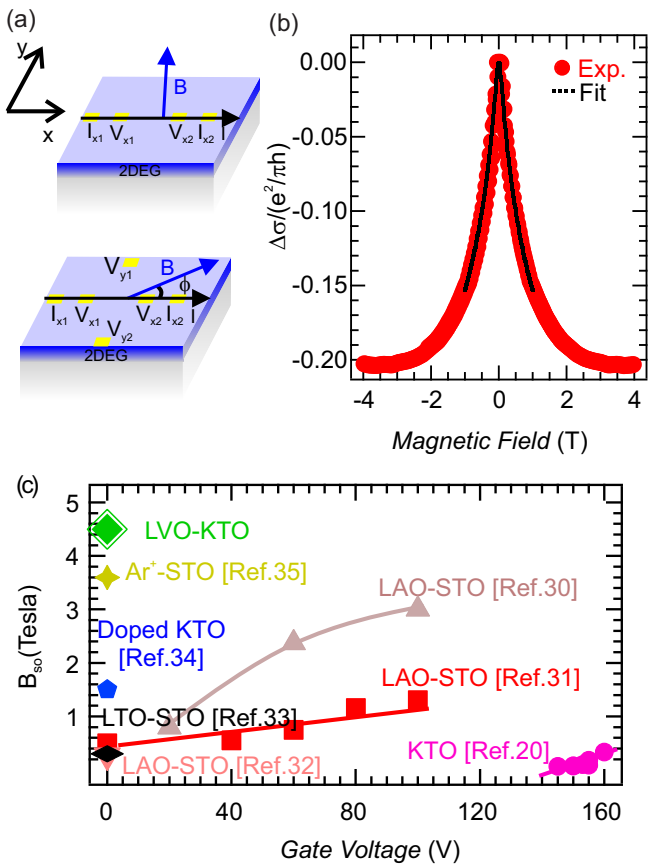}}
\caption{(Color online) (a) (upper panel)  Schematic of the connection geometry for magnetoresistance (R$_{xx}$) measurements for magnetic field applied out-of-plane. (lower panel) for magnetic field applied in the plane. (b) Magnetoconductance plot of 4ml sample as a function of magnetic field showing weak anti-localization due to high spin orbit coupling. (c) Comparitive plot of B$_{SO}$ vs. gate voltage for STO and KTO based systems.}
\end{figure}

\begin{figure*}[t!] \scalebox{2.2}{\includegraphics{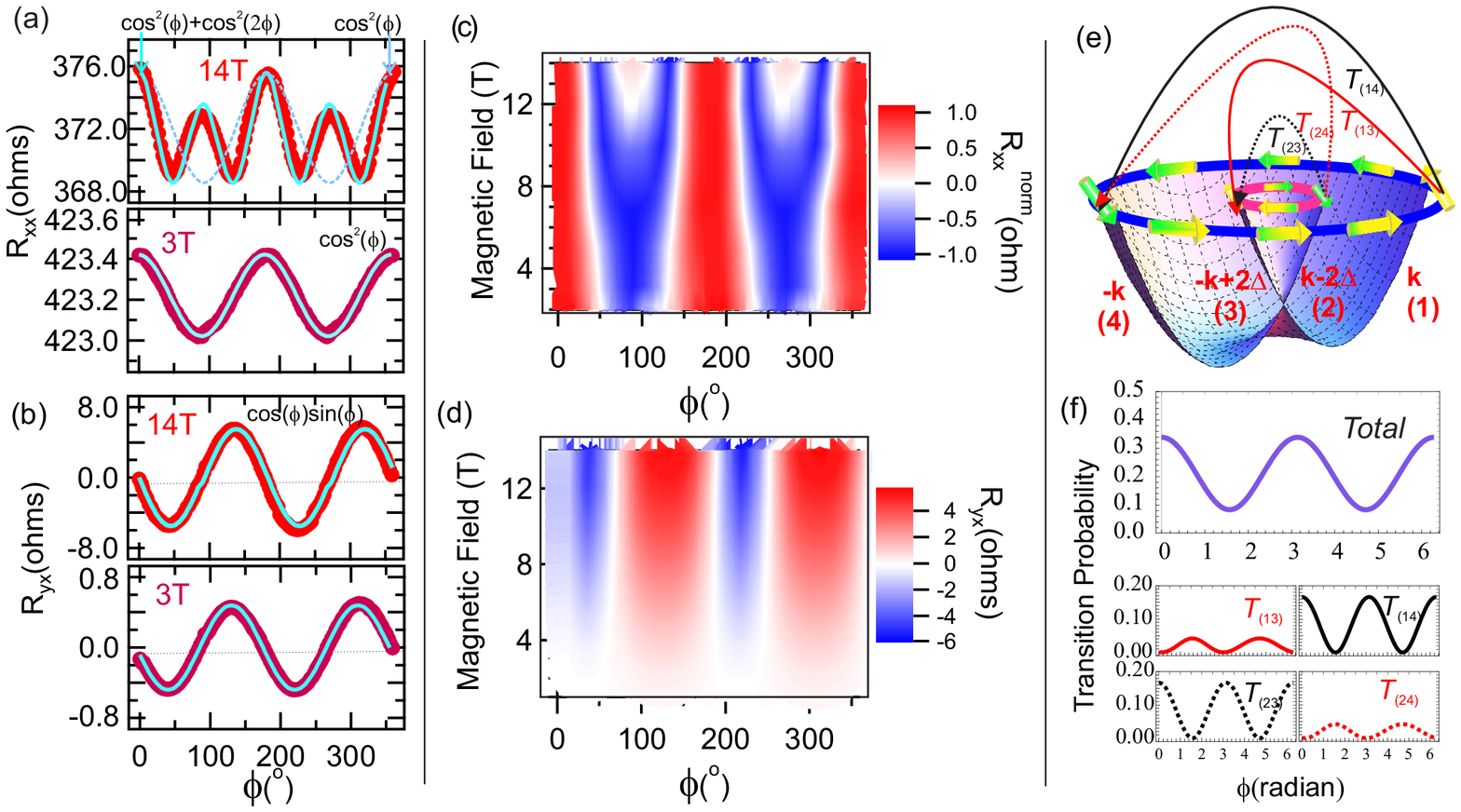}}
\caption{(Color online) (a) and (b) Angle dependent R$_{xx}$ and R$_{yx}$ measured at 1.8 K for 14 T and 3 T. Blue line is the fitted curve. (c) and (d) Applied magnetic field and angle dependent contour plots for normalized R$_{xx}$ and R$_{yx}$. (e) Rahba energy-split bands showing spin-texture at a particular energy and the allowed electronic transitions. (f) Total probabilty and individual probabilities for different allowed electronic transitions between the bands.}
\end{figure*}

Thin films of LVO were grown on (001) oriented Ta-terminated KTO single crystals using pulsed laser deposition (PLD) system. The schematic of the heterostructure is shown in Fig. 1(a). For Ta-termination, method of high temperature annealing followed DI water etching, as reported in our earlier work, was employed.\cite{Tomar2017} Polycrystalline LaVO$_4$ was used as the target material and was ablated using KrF excimer laser at a frequency of  2 Hz. The laser fluence for target ablation was optimized to be 4 Jcm$^{-2}$. During growth, the substrate was kept at an optimized temperature of 600 $^o$C and oxygen partial pressure of the deposition chamber was maintained to be 1x10$^{-6}$ torr.\cite{Hotta2007}  Different samples of varying thickness of LVO were grown. The thickness of the films was controlled using reflection high-energy electron diffraction (RHEED) technique. The RHEED oscillations of the specular spot, for 10 ml, 8 ml, 4 ml sample, as a function of number of unit cells are shown in Fig. 1(b). Figure 1(b) right panel shows the RHEED pattern of the 10 ml sample before and after the film growth. Figure 1(c) shows the XRD plot of the 40 ml sample which confirms the crystalline growth of the LVO film. Inset of Fig. 1(c) shows the rocking curve of the KTO substrate. 

The transport properties of the grown heterostructures were measured using physical property measurement system (PPMS). For temperature dependent resistance measurements, contacts were made by ultrasonically wire-bonding the interface in four probe geometry. Figure 2(a) shows the temperature dependent 2D (r$_{2D}$) and 3D (r$_{3D}$) resistivity for all the samples. The 3 ml sample although conducting at room temperature exhibits an upturn near 30 K. All other samples with LVO more than 3 ml are conducting down to 1.8 K. The 3D resistivity plot shows that the resistivity scales up with thickness confirming that the conductivity comes from the interface only and not the film. Figures 2(b) shows the charge carrier density and mobility (m), for the conducting samples, calculated from the conventional Hall measurements done at 300 K and 1.8 K. It can be seen that above 3 ml of LVO, once the interface becomes conducting, the charge carrier density and mobility are independent of LVO thickness. This is in accordance with the electronic reconstruction mechanism for formation of 2DEG where after achieving the critical thickness to avoid polar catastrophe, increasing the thickness of the film does not add  further carriers at the interface.\cite{Hotta2007} In the present case the critical thickness turns out be 3 ml. We obtained carrier mobility of around 600 cm$^2$V$^{-1}$s$^{-1}$ at 1.8 K in our samples as shown in Fig. 2(b) (lower panel). 

The low temperature conventional magnetoresistance (MR) measurements where magnetic field is applied perpendicular to the interface of LVO and KTO (see Fig. 3(a) upper panel) on 4 ml sample reveals the presence of weak-antilocalization and hence strong spin-orbit coupling in the system.\cite{Nakamura2009,Hikami1980,Iordanskii1994} Theory had been developed by Iordanskii,
Lyanda-Geller, and Pikus (ILP theory) to describe the weak-antilocalization in magnetoconductance for the materials with strong spin-orbit coupling. The expression of the magnetoconductance developed by ILP theory is given by:\cite{Iordanskii1994,Lee2012} 

\begin{equation}
\Delta\sigma= \frac{e^2}{2\pi^2\hbar}[ln(\frac{B_\phi}{B}) - \psi(\frac{1}{2}+\frac{B_{\phi}}{B}) + ln(\frac{B_{SO}}{B}) - \psi(\frac{1}{2} + \frac{B_{SO}}{B})]
\end{equation}

where, B is the applied magnetic field, B$_{\phi}$ ($\hbar$/4el$^2_\phi$) and B$_{SO}$ ($\hbar$/4el$^2_{SO}$) are two characterstic magnetic fields related to phase coherence length (l$_\phi$) and spin-precession length (l$_{SO}$) and $\psi$ is the digamma function. The ILP theory was derived for the magnetic field region B$\textless$ $\hbar$/2el$^2_m$; where l$_m$ is the mean free path of the carriers.\cite{Nakamura2009,Iordanskii1994} For the present sample $\hbar$/2el$^2_m$ is estimated to be 0.3 T. However, we have been able to fit our data upto 1 T. Figure 3(b) shows the magnetoconductance data for 4 ml sample along with the fit using equation 1 (black line).  A high value of B$_{SO}$ $\sim$ 4.4 T  corresponding to a spin-precession length of 6 nm was obtained from the fitting. Phase coherence length of 70 nm and magnetic field strength corresponding to inelasic scattering B$_{\phi}$=0.03 T was obtained for our system. These values of phase coherence length and B$_{\phi}$ are in excellent agrement with the previous report.\cite{Nakamura2009}

In Fig. 3(c), we have plotted the B$_{SO}$ of STO and KTO based systems as a function of applied gate voltage from the literature and compared it with our sample.\cite{Nakamura2009,Caviglia2010,Herranz2015,Gopinadhan2015,Veit2018,King2012,Syro2014} Figure 3(c) clearly suggests that our LVO-KTO interface has the highest B$_{SO}$ among all reported STO and KTO systems.

Figure 3(a) (lower panel) shows a schematic diagram of PHE and AMR measurement configuration, where magnetic field is applied in the sample plane and transverse resistance (R$_{yx}$) and longitudinal resistance (R$_{xx}$) are measured. Usually, PHE and AMR are observed in magnetic systems and are associated with the crystalline symmetry of the system.\cite{Ky1968,Goennenwein2007} Also, recently some topological systems have been reported to witness in-plane AMR and PHE\cite{Taskin2017,Rakhmilevich2018}, the origin of which is anisotropic spin flip transition probabilities arising from broken time reversal symmetery. Theoretically, it has been predicted that the systems with the Rashba-Dressalhaus type of spin band splitting in presence of magnetic impurities may also host in plane AMR and PHE.\cite{Trushin2009,Kozlov2019} Although theoretically predicted, experimental realization of such phenomena in the 2DEG systems with high spin-orbit interaction is not well explored. Considering the large spin-orbit coupling obtained for our system, we expect interesting in-plane AMR and PHE, as well as their evolution as a function of the applied magnetic field.

For these measurements, magnetic field (B) was applied in the sample plane and simultaneous measurements of longitudinal magnetoresistance (R$_{xx}$) and transverse resistance (R$_{yx}$) were made while the varying the angle between I and B. For the first set of experiments, R$_{xx}$ and R$_{yx}$ were measured at 1.8 K by varying the magnitude of applied magnetic field. On scanning the angle between B and I, R$_{xx}$ and R$_{yx}$ were found to show oscillatory behavior. Upto 8 T, we obtained 2 fold periodic oscillations in R$_{xx}$, it slowly changed to 4 fold oscillations above 8 T. Figure 4(a) shows the R$_{xx}$ behavior at 3 T and 14 T. The behavior of normalized R$_{xx}$ on varying the applied magnetic field is shown in the contour plot presented in Fig. 4(c), where R$_{xx}^{norm}$ = (R - R$_{symm}$)/R$_o$. R$_{symm}$ = R$_{min}$ + (R$_{max}$ - R$_{min}$)/2 and  R$_{min}$ is minimum value of R$_{xx}$,  R$_{max}$ is maximum value of  R$_{xx}$ and R$_{o}$ is the value of R$_{xx}$ at 0$^o$.  The low field behavior of R$_{xx}$ is very similar to that observed in topological insulator system of Bi$_{2-x}$Sb$_x$Te$_3$ thin films.\cite{Taskin2017}

We observed oscillations in the planar Hall resistance value as a function of angle between the B and I, with minima at 45$^o$ and maxima at 135$^o$ repeated at 180$^o$ interval. Field dependent measurements were also performed at 1.8 K. Figure 4(d) shows the contour plot of field dependent R$_{yx}$ as a function of angle between B and I at 1.8 K.  It was seen that, on decreasing the magnetic field, the amplitude of oscillations decreases but the nature of oscillations remain same throughout. Figure 4(b) shows the planar hall resistance for 14 T and 3 T field.

The observed two fold oscillations in the resistance on application of an in-plane magnetic field could be intutively understood on the basis of electronic transitions which take place between the Rashba-split energy bands. In LVO-KTO system, due to broken inversion symmetery at the interface and subsequently developed electric field, the relativistic electrons in 5d orbitals of Ta  experience a pseudo magnetic field in the conduction plane and hence may lead to Rashba spin-splitting. The occurrence of a significant spin-orbit interaction has already been reported in the literature, from ARPES measurements in the single crystal of KTO.\cite{King2012}. The presence of a Rashba spin-splitting, that relies on the additional presence of an electric field,  was also seen in this material, for a Fermi wave vector ($\sim$0.2 $\AA^{-1}$ to 0.4 $\AA^{-1}$) at a carrier density of $\sim$2x10$^{14}$ cm$^2$. On the other hand, our system is not just KTO but its interface with LVO (a polar material). Hence, like KTO, the interface, for a (measured) carrier density of 1.02x10$^{14}$ cm$^2$ at a (calculated) Fermi vector of 0.3$\AA^{-1}$, is not only endowed with a non-zero spin-orbit coupling, but is also expected to exhibit a prominent Rashba effect in view of a substantial, polar-polar interface-generated electric field. Our analysis presented below, is based in this premise.  

In our system, the degenerate energy parabola of electrons splits into two parabolas for up-spin and down-spin, generating Rashba spin splitting. Application of an external magnetic field in the conduction plane further adds up a Zeeman splitting term. The external parabola is called the majority band and the internal parabola is called the minority band. Depending on the propagation vector k, spin of the electron, Rashba strength parameter $\alpha$ and the direction and magnitude of the external applied magnetic field, the electrons can make transitions between majority-to-majority (or minority-to-minority) i.e intra-band transitions and majority-to-minority (or minority-to-majority) i.e inter-band transitions. Each allowed transition results in back-scattering of the conduction electrons and hence, contributes to increase in resistance. The energy eigen values for the spin bands can be calculated by solving the hamiltonian\cite{Trushin2009}:
\begin{equation}
H=\epsilon(k)-\alpha(\sigma_xq_y-\sigma_yq_x)
\end{equation}

where,
\begin{equation}
q_x= (-r sin(\Phi) + k_x)
\end{equation}  
\begin{equation}
q_y= (r cos(\Phi) + k_y)
\end{equation} 
 
and r = $\mu_B$ B/$\alpha$. $\epsilon$(k) is free electron energy, $\sigma_(x,y)$ are the Pauli spin matrices, $\Phi$ is the angle between B and I and k$_x$ and k$_y$ are the wave vectors in x and y direction. The electronic transition probability between the bands can be calculated using the eigen vectors for each band and finding the transition matrices. The eigen vectors used for the majority and minority bands are:

\centerline{$\frac{1}{\sqrt{2}}\begin{pmatrix}
1\\
ie^{i\gamma}
\end{pmatrix}$
and $\frac{1}{\sqrt{2}}\begin{pmatrix}
1\\
-ie^{i\gamma}
\end{pmatrix}$}

respectively, where, tan($\gamma$)= (r cos$\Phi$ + k$_y$)/(-r sin$\Phi$ + k$_x$). Figure 4(e) shows the Rashba energy-split bands (numbered as 1,2,3 and 4) with spin texture for a fixed energy value. The allowed transitions between different bands having finite probability are shown with arrows. Because the current is applied along the x-axis (Fig. 3(a)), the relevant momentum component is q$_x$. If we now examine Eq. (2), it is evident that we need to focus only on $\sigma_y$ as far as momentum-reversing transitions (that cause resistance) are concerned. The corresponding transitions matrix elements are decribed in detail in supplementary section.

On application of a magnetic field, intra-band transitions i.e. T$_{14}$ and T$_{23}$ (shown in black arrows) start having an angular dependence and follow a cos$^2$($\Phi$) dependence when the angle between applied magnetic field and current is varied. On the other hand, the transitions presented using red arrows i.e. T$_{13}$ and T$_{24}$ (inter-band transitions) are negligible but start following a sin$^2$($\Phi$) dependence. Therefore, the overall probabilty follows cos$^2$ ($\Phi$) dependence. Figure 4(f) shows individual as well as the total probabilities. Since, each allowed transition results in backscattering of the conduction electrons, the cos$^2$($\Phi$) dependence of the electronic transitions phenomenologically explains the AMR data obtained for low fields as it can be fit completely using cos$^2$($\Phi$) function as shown in Fig. 4(a)(lower panel). At an applied magnetic field of 8 T, we have observed a two to four fold transition in AMR that has cos$^2$($\Phi$)+cos$^2$(2$\Phi$) angular dependence, whereas PHE remains two fold. This 4 fold symmetery of AMR could not be explained using the above phenomenological model. Such two to four fold transitions are observed in STO but these transitions are much complicated and irregular.\cite{Shalom2009,Annadi2013,Ma2017,Rout2017} Such transitions in STO were explained in terms of Liftshitz transitions arising from the  topological change in Fermi surface in presence of intrinsic magnetization of STO. In the present system, we speculate that we might have uncompensated vanadium spins at the interface that could couple to the low dimensional electron gas having relativistic character with strong spin-orbit coupling and give rise to such four fold structure. Our observations suggest a detailed theoretical model of such systems is essential and it would have to contain ingredients of low dimensionality, relativistic electrons, localized magnetic moments and strong spin-orbit coupling.

In conclusion, we have realized a high mobility two dimensional electron gas at a new interface of two polar-polar perovskite oxides. We have observed a high spin-orbit coupling in the system. The magneto-transport measurements show signature of in-plane anisotropic transverse and longitudinal magnetoresistance as a consequence of strong-spin orbit coupling and Rashba spin splitting. The observed nature of the AMR at low magnetic field is phenomenologically understood by using a simplified model with Rashba-spin splitting. The high field four fold AMR warrants an elaborate theoretical analysis. Such a model system may open up an avenue for in depth understanding of the physical properties of low dimensional relativistic electrons in oxide materials with strong spin-orbit coupling. Such detailed understanding might play an important role in the design of new materials for spintronic applications.

SC and NW acknowledge Dr. Denis Maryenko from Centre for Emergent Matter Science (CEMS), RIKEN Japan, for scientific discussions. The financial support from Department of Science and Technology (DST),India - Nano Mission project number (SR/NM/NS-1007/2015) and from Funding Program for World-Leading Innovative R and D on Science and Technology (FIRST) of the Japan Society for the Promotion of Science (JSPS) initiated by the Council for Science and Technology Policy, by JSPS Grants-in Aid for Scientific Research, No. 24226002 is also acknowledged. SD is grateful to the Indian National Science Academy for support through their Senior Scientist scheme and to INST for hospitality.




\begin{thebibliography}{13}
\expandafter\ifx\csname natexlab\endcsname\relax\def\natexlab#1{#1}\fi
\expandafter\ifx\csname bibnamefont\endcsname\relax
  \def\bibnamefont#1{#1}\fi
\expandafter\ifx\csname bibfnamefont\endcsname\relax
  \def\bibfnamefont#1{#1}\fi
\expandafter\ifx\csname citenamefont\endcsname\relax
  \def\citenamefont#1{#1}\fi
\expandafter\ifx\csname url\endcsname\relax
  \def\url#1{\texttt{#1}}\fi
\expandafter\ifx\csname urlprefix\endcsname\relax\def\urlprefix{URL }\fi
\providecommand{\bibinfo}[2]{#2}
\providecommand{\eprint}[2][]{\url{#2}}

\bibitem[{\citenamefont{Datta}(1990)}]{Datta1990}
\bibinfo{author}{\bibfnamefont{S.} \bibnamefont{Datta}},
\bibnamefont{and} {\bibfnamefont{B.} \bibnamefont{Das}},
\bibinfo{journal}{Appl. Phys. Lett.} \textbf{\bibinfo{volume}{56}},
\bibinfo{pages}{665} (\bibinfo{year}{1990}).

\bibitem[{\citenamefont{Meier}(2002)}]{Meier2002}
\bibinfo{author}{\bibfnamefont{G.} \bibnamefont{Meier}},
\bibinfo{author}{\bibfnamefont{T.} \bibnamefont{Matsuyama}},
\bibnamefont{and} {\bibfnamefont{U.} \bibnamefont{Merkt}},
\bibinfo{journal}{Phys. Rev. B} \textbf{\bibinfo{volume}{65}},
\bibinfo{pages}{155322} (\bibinfo{year}{2002}).

\bibitem[{\citenamefont{Bychkov}(1984)}]{Bychkov1984}
\bibinfo{author}{\bibfnamefont{Y.A.} \bibnamefont{Bychkov}},
\bibnamefont{and} {\bibfnamefont{E.I.} \bibnamefont{Rashba}},
\bibinfo{journal}{JETP Lett.} \textbf{\bibinfo{volume}{39}},
\bibinfo{pages}{78} (\bibinfo{year}{1984}).

\bibitem[{\citenamefont{Inoue}(2009)}]{Inoue2009}
\bibinfo{author}{\bibfnamefont{J.I.} \bibnamefont{Inoue}},
\bibinfo{author}{\bibfnamefont{T.} \bibnamefont{Kato}},
\bibinfo{author}{\bibfnamefont{G.E.W.} \bibnamefont{Bauer}},
\bibnamefont{and} {\bibfnamefont{L.W.} \bibnamefont{Molenkamp}},
\bibinfo{journal}{Semicond. Sci.Technol} \textbf{\bibinfo{volume}{24}},
\bibinfo{pages}{064003} (\bibinfo{year}{2009}).

\bibitem[{\citenamefont{Murakawa}(2013)}]{Murakawa2013}
\bibinfo{author}{\bibfnamefont{H.} \bibnamefont{Murakawa}},
\bibinfo{author}{\bibfnamefont{M.S.} \bibnamefont{Bahramy}},
\bibinfo{author}{\bibfnamefont{M.} \bibnamefont{Tokunaga}},
\bibinfo{author}{\bibfnamefont{Y.} \bibnamefont{Kohama}},
\bibinfo{author}{\bibfnamefont{C.} \bibnamefont{Bell}},
\bibinfo{author}{\bibfnamefont{Y.} \bibnamefont{Kaneko}},
\bibinfo{author}{\bibfnamefont{N.} \bibnamefont{Nagaosa}},
\bibinfo{author}{\bibfnamefont{H.} \bibnamefont{Hwang}},
\bibnamefont{and} {\bibfnamefont{Y.} \bibnamefont{Tokura}},
\bibinfo{journal}{Science} \textbf{\bibinfo{volume}{342}},
\bibinfo{pages}{1490} (\bibinfo{year}{2013}).

\bibitem[{\citenamefont{Eisentein}(1984)}]{Eisentein1984}
\bibinfo{author}{\bibfnamefont{J.P.} \bibnamefont{Eisentein}},
\bibinfo{author}{\bibfnamefont{H.L.} \bibnamefont{Stormer}},
\bibinfo{author}{\bibfnamefont{V.} \bibnamefont{Narayanamurti}},
\bibinfo{author}{\bibfnamefont{A.C.} \bibnamefont{Gossard}},
\bibnamefont{and} {\bibfnamefont{W.} \bibnamefont{Wiegmann}},
\bibinfo{journal}{Phys. Rev. Lett.} \textbf{\bibinfo{volume}{53}},
\bibinfo{pages}{2579} (\bibinfo{year}{1984}).

\bibitem[{\citenamefont{Nitta}(1997)}]{Nitta1997}
\bibinfo{author}{\bibfnamefont{J.} \bibnamefont{Nitta}},
\bibinfo{author}{\bibfnamefont{H.} \bibnamefont{Takayanagi}},
\bibnamefont{and} {\bibfnamefont{T.} \bibnamefont{Enoki}},
\bibinfo{journal}{Phys. Rev. Lett.} \textbf{\bibinfo{volume}{78}},
\bibinfo{pages}{1335} (\bibinfo{year}{1997}).

\bibitem[{\citenamefont{Bibes}(2011)}]{Bibes2011}
\bibinfo{author}{\bibfnamefont{M.} \bibnamefont{Bibes}},
\bibinfo{author}{\bibfnamefont{J.E.} \bibnamefont{Villegas}},
\bibnamefont{and} {\bibfnamefont{A.} \bibnamefont{Barthelemy}},
\bibinfo{journal}{Advances in Physics} \textbf{\bibinfo{volume}{60(1)}},
\bibinfo{pages}{5} (\bibinfo{year}{2011}).

\bibitem[{\citenamefont{Shanavas}(2014)}]{Shanavas2014}
\bibinfo{author}{\bibfnamefont{K.V.} \bibnamefont{Shanavas}},
\bibinfo{author}{\bibfnamefont{Z.S.} \bibnamefont{Popovic}},
\bibnamefont{and} {\bibfnamefont{S.} \bibnamefont{Satpathy}},
\bibinfo{journal}{Phys. Rev. B} \textbf{\bibinfo{volume}{90}},
\bibinfo{pages}{165108} (\bibinfo{year}{2014}).

\bibitem[{\citenamefont{Tokura}(1994)}]{Tokura1994}
\bibinfo{author}{\bibfnamefont{Y.} \bibnamefont{Tokura}},
\bibinfo{author}{\bibfnamefont{A.} \bibnamefont{Urushibara}},
\bibinfo{author}{\bibfnamefont{Y.} \bibnamefont{Moritomo}},
\bibinfo{author}{\bibfnamefont{T.} \bibnamefont{Arima}},
\bibinfo{author}{\bibfnamefont{A.} \bibnamefont{Asamitsu}},
\bibinfo{author}{\bibfnamefont{G.} \bibnamefont{Kido}},
\bibnamefont{and} {\bibfnamefont{N.} \bibnamefont{Furukawa}},
\bibinfo{journal}{J. Phys. Soc. Jpn.} \textbf{\bibinfo{volume}{63}},
\bibinfo{pages}{3931} (\bibinfo{year}{1994}).

\bibitem[{\citenamefont{Schooley}(1964)}]{Schooley1964}
\bibinfo{author}{\bibfnamefont{J.F.} \bibnamefont{Schooley}},
\bibinfo{author}{\bibfnamefont{W.R.} \bibnamefont{Hosler}},
\bibnamefont{and} {\bibfnamefont{M.L.} \bibnamefont{Cohen}},
\bibinfo{journal}{Phys. Rev. Lett.} \textbf{\bibinfo{volume}{17}},
\bibinfo{pages}{474} (\bibinfo{year}{1964}).

\bibitem[{\citenamefont{Tikhomirov}(2002)}]{Tikhomirov2002}
\bibinfo{author}{\bibfnamefont{O.} \bibnamefont{Tikhomirov}}
\bibinfo{author}{\bibfnamefont{H.} \bibnamefont{Jiang}}
\bibnamefont{and} {\bibfnamefont{J.} \bibnamefont{Levy}},
\bibinfo{journal}{Phys. Rev. Lett.} \textbf{\bibinfo{volume}{89}},
\bibinfo{pages}{147601} (\bibinfo{year}{2002}).

\bibitem[{\citenamefont{Lee}(2013)}]{Lee2013}
\bibinfo{author}{\bibfnamefont{J.S.} \bibnamefont{Lee}},
\bibinfo{author}{\bibfnamefont{Y.W.} \bibnamefont{Xie}},
\bibinfo{author}{\bibfnamefont{H.K.} \bibnamefont{Sato}},
\bibinfo{author}{\bibfnamefont{C.} \bibnamefont{Bell}},
\bibinfo{author}{\bibfnamefont{Y.} \bibnamefont{Hikita}},
\bibinfo{author}{\bibfnamefont{H.Y.} \bibnamefont{Hwang}},
\bibnamefont{and} {\bibfnamefont{C.C.} \bibnamefont{Kao}},
\bibinfo{journal}{Nat. Mater.} \textbf{\bibinfo{volume}{12}},
\bibinfo{pages}{703} (\bibinfo{year}{2013}).



\bibitem[{\citenamefont{Ohtomo}(2004)}]{Ohtomo2004}
\bibinfo{author}{\bibfnamefont{A.} \bibnamefont{Ohtomo}}
\bibnamefont{and} {\bibfnamefont{H.Y.} \bibnamefont{Hwang}},
\bibinfo{journal}{Nature} \textbf{\bibinfo{volume}{47}},
\bibinfo{pages}{424} (\bibinfo{year}{2004}).

\bibitem[{\citenamefont{Hotta}(2007)}]{Hotta2007}
\bibinfo{author}{\bibfnamefont{Y.} \bibnamefont{Hotta}},
\bibinfo{author}{\bibfnamefont{T.} \bibnamefont{Susaki}},
\bibnamefont{and} {\bibfnamefont{H.Y.} \bibnamefont{Hwang}},
\bibinfo{journal}{Phys. Rev. Lett.} \textbf{\bibinfo{volume}{99}},
\bibinfo{pages}{236805} (\bibinfo{year}{2007}).


\bibitem[{\citenamefont{Chen}(2015)}]{Chen2015}
\bibinfo{author}{\bibfnamefont{Y.} \bibnamefont{Chen}},
\bibinfo{author}{\bibfnamefont{F.} \bibnamefont{Trier}},
\bibinfo{author}{\bibfnamefont{T.} \bibnamefont{Kasama}},
\bibinfo{author}{\bibfnamefont{D.V.} \bibnamefont{Christenen}},
\bibinfo{author}{\bibfnamefont{N.} \bibnamefont{Bovet}},
\bibinfo{author}{\bibfnamefont{Z.I.} \bibnamefont{Balogh}},
\bibinfo{author}{\bibfnamefont{H.} \bibnamefont{Li}},
\bibinfo{author}{\bibfnamefont{K.T.S.} \bibnamefont{Thyden}},
\bibinfo{author}{\bibfnamefont{W.} \bibnamefont{Zhang}},
\bibinfo{author}{\bibfnamefont{S.} \bibnamefont{Yazdi}},
\bibinfo{author}{\bibfnamefont{P.} \bibnamefont{Norby}},
\bibinfo{author}{\bibfnamefont{N.} \bibnamefont{Pryds}},
\bibnamefont{and} {\bibfnamefont{S.} \bibnamefont{Linderoth}},
\bibinfo{journal}{Nano Lett.} \textbf{\bibinfo{volume}{15}},
\bibinfo{pages}{1849} (\bibinfo{year}{2015}).

\bibitem[{\citenamefont{Tomar}(2017)}]{Tomar2017}
\bibinfo{author}{\bibfnamefont{R.} \bibnamefont{Tomar}},
\bibinfo{author}{\bibfnamefont{N.} \bibnamefont{Wadehra}},
\bibinfo{author}{\bibfnamefont{V.} \bibnamefont{Budhiraja}},
\bibinfo{author}{\bibfnamefont{B.} \bibnamefont{Prakash}},
\bibnamefont{and} {\bibfnamefont{S.} \bibnamefont{Chakraverty}},
\bibinfo{journal}{Appl. Surf. Sci.} \textbf{\bibinfo{volume}{427}},
\bibinfo{pages}{861} (\bibinfo{year}{2017}).

\bibitem[{\citenamefont{Shanavas}(2014)}]{Shanavas2014a}
\bibinfo{author}{\bibfnamefont{K.V.} \bibnamefont{Shanavas}},
\bibnamefont{and} {\bibfnamefont{S.} \bibnamefont{Satpathy}},
\bibinfo{journal}{Phys. Rev. Lett.} \textbf{\bibinfo{volume}{112}},
\bibinfo{pages}{086802} (\bibinfo{year}{2014}).

\bibitem[{\citenamefont{Wadehra}(2017)}]{Wadehra2017}
\bibinfo{author}{\bibfnamefont{N.} \bibnamefont{Wadehra}},
\bibinfo{author}{\bibfnamefont{R.} \bibnamefont{Tomar}},
\bibinfo{author}{\bibfnamefont{S.} \bibnamefont{Halder}},
\bibinfo{author}{\bibfnamefont{M.} \bibnamefont{Sharma}},
\bibinfo{author}{\bibfnamefont{I.} \bibnamefont{Singh}},
\bibinfo{author}{\bibfnamefont{N.} \bibnamefont{Jena}},
\bibinfo{author}{\bibfnamefont{B.} \bibnamefont{Prakash}},
\bibinfo{author}{\bibfnamefont{A.D} \bibnamefont{Sarkar}},
\bibinfo{author}{\bibfnamefont{C.} \bibnamefont{Bera}},
\bibinfo{author}{\bibfnamefont{A.} \bibnamefont{Venkatesan}},
\bibnamefont{and} {\bibfnamefont{S.} \bibnamefont{Chakraverty}},
\bibinfo{journal}{Phys. Rev. B} \textbf{\bibinfo{volume}{96}},
\bibinfo{pages}{115423} (\bibinfo{year}{2017}).



\bibitem[{\citenamefont{Nakamura}(2009)}]{Nakamura2009}
\bibinfo{author}{\bibfnamefont{H.} \bibnamefont{Nakamura}},
\bibnamefont{and} {\bibfnamefont{T.} \bibnamefont{Kimura}},
\bibinfo{journal}{Phys. Rev. B} \textbf{\bibinfo{volume}{80}},
\bibinfo{pages}{121308} (\bibinfo{year}{2009}).

\bibitem[{\citenamefont{Ky}(1968)}]{Ky1968}
\bibinfo{author}{\bibfnamefont{V.D.} \bibnamefont{Ky}}
\bibinfo{journal}{Phys. Stat. Sol.} \textbf{\bibinfo{volume}{26}},
\bibinfo{pages}{565} (\bibinfo{year}{1968}).


\bibitem[{\citenamefont{Goennenwein}(2007)}]{Goennenwein2007}
\bibinfo{author}{\bibfnamefont{S.T.B.} \bibnamefont{Goennenwein}},
\bibinfo{author}{\bibfnamefont{R.S.} \bibnamefont{Keizer}},
\bibinfo{author}{\bibfnamefont{S.W.} \bibnamefont{Schink}},
\bibinfo{author}{\bibfnamefont{I.V} \bibnamefont{Dijk}},
\bibinfo{author}{\bibfnamefont{T.M..} \bibnamefont{Klapwijk}},
\bibinfo{author}{\bibfnamefont{G.X.} \bibnamefont{Miao}},
\bibinfo{author}{\bibfnamefont{G.} \bibnamefont{Xiao}},
\bibnamefont{and} {\bibfnamefont{A.} \bibnamefont{Gupta}},
\bibinfo{journal}{Appl. Phys. Lett.} \textbf{\bibinfo{volume}{90}},
\bibinfo{pages}{90} (\bibinfo{year}{2007}).

\bibitem[{\citenamefont{Taskin}(2017)}]{Taskin2017}
\bibinfo{author}{\bibfnamefont{A.A.} \bibnamefont{Taskin}},
\bibnamefont{author} {\bibfnamefont{H.F.} \bibnamefont{Legg}},
\bibinfo{author}{\bibfnamefont{F.} \bibnamefont{Yang}},
\bibinfo{author}{\bibfnamefont{S.} \bibnamefont{Sasaki}},
\bibinfo{author}{\bibfnamefont{Y.} \bibnamefont{Kanai}},
\bibinfo{author}{\bibfnamefont{K.} \bibnamefont{Matsumoto}},
\bibinfo{author}{\bibfnamefont{A.} \bibnamefont{Rosch}},
\bibnamefont{and} {\bibfnamefont{Y.} \bibnamefont{Ando}},
\bibinfo{journal}{Nat. Commun.} \textbf{\bibinfo{volume}{8}},
\bibinfo{pages}{1340-1} (\bibinfo{year}{2017}).

\bibitem[{\citenamefont{Rakhmilevich}(2018)}]{Rakhmilevich2018}
\bibinfo{author}{\bibfnamefont{D.} \bibnamefont{Rakhmilevich}},
\bibinfo{author}{\bibfnamefont{F.} \bibnamefont{Wang}},
\bibinfo{author}{\bibfnamefont{W.} \bibnamefont{Zhao}},
\bibinfo{author}{\bibfnamefont{M.H.} \bibnamefont{Chan}},
\bibinfo{author}{\bibfnamefont{J.S.} \bibnamefont{Moodera}},
\bibinfo{author}{\bibfnamefont{C.} \bibnamefont{Liu}},
\bibnamefont{and} {\bibfnamefont{C.Z} \bibnamefont{Chang}},
\bibinfo{journal}{Phys. Rev. B} \textbf{\bibinfo{volume}{98}},
\bibinfo{pages}{094404} (\bibinfo{year}{2018}).

\bibitem[{\citenamefont{Trushin}(2009)}]{Trushin2009}
\bibinfo{author}{\bibfnamefont{M.} \bibnamefont{Trushin}},
\bibinfo{author}{\bibfnamefont{K.} \bibnamefont{Vyborny}},
\bibinfo{author}{\bibfnamefont{P.} \bibnamefont{Moraczewski}},
\bibinfo{author}{\bibfnamefont{A.A.} \bibnamefont{Kovalev}},
\bibinfo{author}{\bibfnamefont{J.} \bibnamefont{Schliemann}},
\bibnamefont{and} {\bibfnamefont{T.} \bibnamefont{Jungwirth}},
\bibinfo{journal}{Phys. Rev. B} \textbf{\bibinfo{volume}{80}},
\bibinfo{pages}{134405} (\bibinfo{year}{2009}).

\bibitem[{\citenamefont{Kozlov}(2019)}]{Kozlov2019}
\bibinfo{author}{\bibfnamefont{I.V.} \bibnamefont{Kozlov}},
\bibnamefont{and} {\bibfnamefont{Y.A.} \bibnamefont{Kolesnichenko}},
\bibinfo{journal}{Phys. Rev. B} \textbf{\bibinfo{volume}{99}},
\bibinfo{pages}{085129} (\bibinfo{year}{2019}).

\bibitem[{\citenamefont{Hikami}(1980)}]{Hikami1980}
\bibinfo{author}{\bibfnamefont{S.} \bibnamefont{Hikami}},
\bibinfo{author}{\bibfnamefont{A.I.} \bibnamefont{Larkin}},
\bibnamefont{and} {\bibfnamefont{Y.} \bibnamefont{Nagaoka}},
\bibinfo{journal}{Prog. Theor. Phys.} \textbf{\bibinfo{volume}{63}},
\bibinfo{pages}{707} (\bibinfo{year}{1980}).

\bibitem[{\citenamefont{Iordanskii}(1994)}]{Iordanskii1994}
\bibinfo{author}{\bibfnamefont{S.V.} \bibnamefont{Iordanskii}},
\bibinfo{author}{\bibfnamefont{Y.B.} \bibnamefont{Lyanda-Geller}},
\bibnamefont{and} {\bibfnamefont{G.E.} \bibnamefont{Pikus}},
\bibinfo{journal}{JETP Lett.} \textbf{\bibinfo{volume}{60}},
\bibinfo{pages}{206} (\bibinfo{year}{1994}).

\bibitem[{\citenamefont{Lee}(2012)}]{Lee2012}
\bibinfo{author}{\bibfnamefont{J.} \bibnamefont{Lee}},
\bibinfo{author}{\bibfnamefont{J.} \bibnamefont{Park}},
\bibinfo{author}{\bibfnamefont{J.H.} \bibnamefont{Lee}},
\bibinfo{author}{\bibfnamefont{J.S.} \bibnamefont{Kim}},
\bibnamefont{and} {\bibfnamefont{H.J.} \bibnamefont{Lee}},
\bibinfo{journal}{Phys. Rev. B} \textbf{\bibinfo{volume}{86}},
\bibinfo{pages}{245321} (\bibinfo{year}{2012}).

\bibitem[{\citenamefont{Caviglia}(2010)}]{Caviglia2010}
\bibinfo{author}{\bibfnamefont{A.D.} \bibnamefont{Caviglia}},
\bibinfo{author}{\bibfnamefont{M.} \bibnamefont{Gabay}},
\bibinfo{author}{\bibfnamefont{S.} \bibnamefont{Gariglio}},
\bibinfo{author}{\bibfnamefont{N.} \bibnamefont{Reyren}},
\bibinfo{author}{\bibfnamefont{C.} \bibnamefont{Cancellieri}},
\bibnamefont{and} {\bibfnamefont{J.M.} \bibnamefont{Triscone}},
\bibinfo{journal}{Phys. Rev. Lett.} \textbf{\bibinfo{volume}{104}},
\bibinfo{pages}{126803} (\bibinfo{year}{2010}).

\bibitem[{\citenamefont{Herranz}(2015)}]{Herranz2015}
\bibinfo{author}{\bibfnamefont{G.} \bibnamefont{Herranz}},
\bibinfo{author}{\bibfnamefont{G.} \bibnamefont{Singh}},
\bibinfo{author}{\bibfnamefont{N.} \bibnamefont{Bergeal}},
\bibinfo{author}{\bibfnamefont{A.} \bibnamefont{Jouan}},
\bibinfo{author}{\bibfnamefont{J.} \bibnamefont{Lesueur}},
\bibinfo{author}{\bibfnamefont{J.} \bibnamefont{Gazquez}},
\bibinfo{author}{\bibfnamefont{M.} \bibnamefont{Varela}},
\bibinfo{author}{\bibfnamefont{M.} \bibnamefont{Scigaj}},
\bibinfo{author}{\bibfnamefont{N.} \bibnamefont{Dix}},
\bibinfo{author}{\bibfnamefont{F.} \bibnamefont{Sanchez}},
\bibnamefont{and} {\bibfnamefont{J.} \bibnamefont{Fontcuberta}},
\bibinfo{journal}{Nat. Commun.} \textbf{\bibinfo{volume}{6}},
\bibinfo{pages}{6028} (\bibinfo{year}{2015}).

\bibitem[{\citenamefont{Gopinadhan}(2015)}]{Gopinadhan2015}
\bibinfo{author}{\bibfnamefont{K.} \bibnamefont{Gopinadhan}},
\bibinfo{author}{\bibfnamefont{A.} \bibnamefont{Annadi}},
\bibinfo{author}{\bibfnamefont{Y.} \bibnamefont{Kim}},
\bibinfo{author}{\bibfnamefont{A.} \bibnamefont{Srivastava}},
\bibinfo{author}{\bibfnamefont{B.} \bibnamefont{Kumar}},
\bibinfo{author}{\bibfnamefont{J.} \bibnamefont{Chen}},
\bibinfo{author}{\bibfnamefont{J.M.D.} \bibnamefont{Coey}},
\bibinfo{author}{\bibfnamefont{} \bibnamefont{Ariando}},
\bibnamefont{and} {\bibfnamefont{T.} \bibnamefont{Venkatesan}},
\bibinfo{journal}{Adv. Electron. Mater.} \textbf{\bibinfo{volume}{1}},
\bibinfo{pages}{1500114} (\bibinfo{year}{2015}).

\bibitem[{\citenamefont{Veit}(2018)}]{Veit2018}
\bibinfo{author}{\bibfnamefont{M.J.} \bibnamefont{Veit}},
\bibinfo{author}{\bibfnamefont{R.} \bibnamefont{Arras}},
\bibinfo{author}{\bibfnamefont{B.J.} \bibnamefont{Ramshaw}},
\bibinfo{author}{\bibfnamefont{R.} \bibnamefont{Pentcheva}},
\bibnamefont{and} {\bibfnamefont{Y.} \bibnamefont{Suzuki}},
\bibinfo{journal}{Nat. Commun.} \textbf{\bibinfo{volume}{9}},
\bibinfo{pages}{1458} (\bibinfo{year}{2018}).

\bibitem[{\citenamefont{King}(2012)}]{King2012}
\bibinfo{author}{\bibfnamefont{P.D.C.} \bibnamefont{King}},
\bibinfo{author}{\bibfnamefont{R.H.} \bibnamefont{He}},
\bibinfo{author}{\bibfnamefont{T.} \bibnamefont{Eknapakul}},
\bibinfo{author}{\bibfnamefont{S.K.} \bibnamefont{Mo}},
\bibinfo{author}{\bibfnamefont{Y.} \bibnamefont{Kaneko}},
\bibinfo{author}{\bibfnamefont{S.} \bibnamefont{Harashima}},
\bibinfo{author}{\bibfnamefont{Y.} \bibnamefont{Hikita}},
\bibinfo{author}{\bibfnamefont{M.S.} \bibnamefont{Bahramy}},
\bibinfo{author}{\bibfnamefont{C.} \bibnamefont{Bell}},
\bibinfo{author}{\bibfnamefont{Z.} \bibnamefont{Hussain}},
\bibinfo{author}{\bibfnamefont{Y.} \bibnamefont{Tokura}},
\bibinfo{author}{\bibfnamefont{Z.X.} \bibnamefont{Shen}},
\bibinfo{author}{\bibfnamefont{H.Y.} \bibnamefont{Hwang}},
\bibinfo{author}{\bibfnamefont{F.} \bibnamefont{Baumberger}},
\bibnamefont{and} {\bibfnamefont{W.} \bibnamefont{Meevasana}},
\bibinfo{journal}{Phys. Rev. Lett.} \textbf{\bibinfo{volume}{108}},
\bibinfo{pages}{117602} (\bibinfo{year}{2012}).

\bibitem[{\citenamefont{Syro}(2014)}]{Syro2014}
\bibinfo{author}{\bibfnamefont{A.F.S.} \bibnamefont{Syro}},
\bibinfo{author}{\bibfnamefont{F.} \bibnamefont{Fortuna}},
\bibinfo{author}{\bibfnamefont{C.} \bibnamefont{Bareille}},
\bibinfo{author}{\bibfnamefont{T.C.} \bibnamefont{Rodel}},
\bibinfo{author}{\bibfnamefont{G.} \bibnamefont{Landolt}},
\bibinfo{author}{\bibfnamefont{N.C.} \bibnamefont{Plumb}},
\bibinfo{author}{\bibfnamefont{J.H.} \bibnamefont{Dil}},
\bibnamefont{and} {\bibfnamefont{M.} \bibnamefont{Radovic}},
\bibinfo{journal}{Nat. Mater.} \textbf{\bibinfo{volume}{13}},
\bibinfo{pages}{1085} (\bibinfo{year}{2014}).

\bibitem[{\citenamefont{Shalom}(2009)}]{Shalom2009}
\bibinfo{author}{\bibfnamefont{M.B.} \bibnamefont{Shalom}},
\bibinfo{author}{\bibfnamefont{C.W.} \bibnamefont{Tai}},
\bibinfo{author}{\bibfnamefont{Y.} \bibnamefont{Lereah}},
\bibinfo{author}{\bibfnamefont{M.} \bibnamefont{Sachs}},
\bibinfo{author}{\bibfnamefont{E.} \bibnamefont{Levy}},
\bibinfo{author}{\bibfnamefont{D.} \bibnamefont{Rakhmilevitch}},
\bibinfo{author}{\bibfnamefont{A.} \bibnamefont{Palevski}},
\bibnamefont{and} {\bibfnamefont{Y.} \bibnamefont{Dagan}},
\bibinfo{journal}{Phys. Rev. B} \textbf{\bibinfo{volume}{80}},
\bibinfo{pages}{140403} (\bibinfo{year}{2009}).

\bibitem[{\citenamefont{Annadi}(2013)}]{Annadi2013}
\bibinfo{author}{\bibfnamefont{A.} \bibnamefont{Annadi}},
\bibinfo{author}{\bibfnamefont{Z.} \bibnamefont{Huang}},
\bibinfo{author}{\bibfnamefont{K.} \bibnamefont{Gopinadhan}},
\bibinfo{author}{\bibfnamefont{X.R.} \bibnamefont{Wang}},
\bibinfo{author}{\bibfnamefont{A.} \bibnamefont{Srivastava}},
\bibinfo{author}{\bibfnamefont{Z.Q.} \bibnamefont{Liu}},
\bibinfo{author}{\bibfnamefont{L.C.} \bibnamefont{Zhang}},
\bibinfo{author}{\bibfnamefont{H.H} \bibnamefont{Ma}},
\bibinfo{author}{\bibfnamefont{T.P.} \bibnamefont{Sarkar}},
\bibinfo{author}{\bibfnamefont{T.} \bibnamefont{Venkatesan}},
\bibnamefont{and} {\bibfnamefont{} \bibnamefont{Ariando}},
\bibinfo{journal}{Phys. Rev. B} \textbf{\bibinfo{volume}{87}},
\bibinfo{pages}{201102(R)} (\bibinfo{year}{2013}).

\bibitem[{\citenamefont{Ma}(2017)}]{Ma2017}
\bibinfo{author}{\bibfnamefont{H.J.H.} \bibnamefont{Ma}},
\bibinfo{author}{\bibfnamefont{J.} \bibnamefont{Zhou}},
\bibinfo{author}{\bibfnamefont{M.} \bibnamefont{Yang}},
\bibinfo{author}{\bibfnamefont{Y.} \bibnamefont{Liu}},
\bibinfo{author}{\bibfnamefont{S.W.} \bibnamefont{Zeng}},
\bibinfo{author}{\bibfnamefont{W.X.} \bibnamefont{Zhou}},
\bibinfo{author}{\bibfnamefont{L.C.} \bibnamefont{Zhang}},
\bibinfo{author}{\bibfnamefont{T.} \bibnamefont{Venkatesan}},
\bibinfo{author}{\bibfnamefont{Y.P.} \bibnamefont{Feng}},
\bibnamefont{and} {\bibfnamefont{} \bibnamefont{Ariando}},
\bibinfo{journal}{Phys. Rev. B} \textbf{\bibinfo{volume}{95}},
\bibinfo{pages}{155314} (\bibinfo{year}{2017}).

\bibitem[{\citenamefont{Rout}(2017)}]{Rout2017}
\bibinfo{author}{\bibfnamefont{P.K.} \bibnamefont{Rout}},
\bibinfo{author}{\bibfnamefont{I.} \bibnamefont{Agireen}},
\bibinfo{author}{\bibfnamefont{E.} \bibnamefont{Maniv}},
\bibinfo{author}{\bibfnamefont{M.} \bibnamefont{Goldstein}},
\bibnamefont{and} {\bibfnamefont{Y} \bibnamefont{Dagan}},
\bibinfo{journal}{Phys. Rev. B} \textbf{\bibinfo{volume}{95}},
\bibinfo{pages}{241107(R)} (\bibinfo{year}{2017}).




\end{thebibliography}
\end{document}